\documentclass{article}

\usepackage{latexsym}
\usepackage{amsmath}
\usepackage{amsthm}
\usepackage{amssymb}
\usepackage{stmaryrd}
\usepackage{xspace}
\usepackage{graphics}
\usepackage{diagrams}  

\newcommand{\figureline}{\rule{\textwidth}{0.5pt}}
\newcommand{\figureend}{\rule{\textwidth}{0.5pt}}

\newtheorem{theorem}{Theorem}
\newtheorem*{theorem*}{Theorem}
\newtheorem{lemma}[theorem]{Lemma}

\newtheorem{proposition}[theorem]{Proposition}
\newtheorem{definition}[theorem]{Definition}

\newtheorem{example}[theorem]{Example}
\newtheorem{remark}{Remark}

\newcommand{\denote}[1]{\llbracket #1 \rrbracket} 
\newcommand{\name}[1]{\ulcorner #1 \urcorner}
\newcommand{\coname}[1]{\llcorner #1 \lrcorner}
\newcommand{\iso}{\cong}
\newcommand{\isomorphism}{\cong}

\newcommand{\ds}{^{\prime}} 
\newcommand{\dagg}{^\dagger} 

\newcommand{\catA}{\ensuremath{{\cal A}}\xspace}

\newcommand{\catC}{\ensuremath{{\cal C}}\xspace}

\newcommand{\fdhilb}{\ensuremath{\textbf{FDHilb}}\xspace}

\newcommand{\catRel}{\ensuremath{\textbf{Rel}}\xspace}
\newcommand{\rel}{\ensuremath{\textbf{Rel}}\xspace}

\newcommand{\id}[1]{\ensuremath{1_{#1}}}

\newcommand{\ie}{\textit{i.e.}~}
\newcommand{\InvCat}{\mathbf{InvCat}}
\newcommand{\CC}{\mathcal{C}}
\newcommand{\rarr}{\rightarrow}

\newcommand{\ket}[1]{
    \ensuremath{\left|  #1 \right\rangle}\xspace}

\newcommand{\innp}[2]{
    \ensuremath{\langle #1 \mid #2 \rangle}}

\newcommand{\ZERO}{\mathbf{0}}
\newcommand{\I}{\mathrm{I}}

\begin{document}


\title{A Categorical Quantum Logic}
\author{Samson Abramsky \;\;\; Ross Duncan \\\\
  Oxford University Computing Laboratory}

\date{15 December 2005}

\maketitle

\begin{abstract}
  We define a strongly normalising proof-net calculus corresponding to the logic
  of strongly compact closed categories with biproducts.  The calculus
  is a full and faithful representation of the free strongly compact
  closed category with biproducts on a given category with an involution.  This syntax
  can be used to represent and reason about quantum processes.
\end{abstract}

\section{Introduction}
Recent work by Abramsky and Coecke \cite{AbrCoe:CatSemQuant:2004}
develops a complete axiomatization of finite dimensional quantum
mechanics in the abstract setting of strongly compact closed
categories with biproducts. This is used to formalize and verify a
number of key quantum information protocols.  In this setting,
classical information flow is explicitly represented by the biproduct
structure, while the compact closed structure models quantum
behaviour: preparation, unitary evolution and projection, including
powerful algebraic methods for representing and reasoning about
entangled states.  Mediating between the two levels is a semiring of
scalars, which is an intrinsic part of the structure, and represents
the probability amplitudes in the abstract.

Compact closed categories can be seen as degenerate models of
multiplicative linear logic in which the connectives, tensor and par,
are identified.  Similarly, the biproduct is a connective in which the
additives of linear logic are combined.  The system resulting from
these identifications has a very different flavour to linear logic,
indeed to any familiar system of logic.  Cyclic structures abound, and
every sequent is provable. These apparent perversities are, however,
no cause for alarm: the resulting equations faithfully mirror
calculations in quantum mechanics as shown in
\cite{AbrCoe:CatSemQuant:2004}. Moreover, the cyclic proof structures
give rise to \emph{scalars}, and allow quantitative aspects to be
expressed.

Beginning with a category \catA of basic types and maps between them
we develop a logical presentation of the free strongly compact closed
category with biproducts $F\catA$.  By varying the choice of \catA it
is possible to explore what are the minimal requirements to achieve
various ``quantum'' effects.  For example, let \catA be the category
with the single object $\mathbb{C}^2$, and the Pauli maps as the
non-identity arrows.  In this case the only possible preparations are
the elements of the Bell basis, and their composites. This is
sufficient for entanglement swapping, but not logic gate
teleportation.

We extend the work of Kelly and Laplaza \cite{KelLap:comcl:1980} by
explicitly describing the arrows of this category in terms of a system
of proof-nets.  We prove that this syntax is a faithful and fully
complete representation of $F\catA$.

In this system the axiom links represent the preparation of atomic
states, while cuts encode projections.  The biproduct is used to
represent the classical branching structure.  Hence proof-nets can
encode physical networks of quantum state preparations and
measurements, and a compilation process to quantum circuits is easily
defined.  Cut-elimination reduces each such network to one without
measurements and as such expresses the outcome of executing  a quantum protocol or
algorithm.  Cut-elimination preserves denotational equality and hence
can serve as a correctness proof for the protocol encoded by the
proof-net. 

The cut-elimination procedure provides an easily implemented method
for performing calculations about the structure of entangled
states. Such a concrete implementation promises to be a useful tool
for reasoning qualitatively about quantum protocols.  For example, in
\cite{AbrCoe:CatSemQuant:2004}, the correctness of several significant quantum protocols is captured as the commutativity of a certain diagram, which expresses the fact that the protocol meets its specification. Such reasoning can  be automated by representing the protocol as a proof-net containing Cuts, and normalizing it to show its equality with the specification, which can be represented as a Cut-free proof-net.

Furthermore,  the
system can be viewed as a step towards a quantum programming language
equipped with an entanglement-aware type system.

In order to give a flavour of how the proof-net syntax may be used to
represent quantum systems we develop an example, the entanglement
swapping protocol, in the next section.  In
section~\ref{sec:categ-prel} we reprise the requisite categorical
structures, and in sections~\ref{sec:logic-scccbs} and \ref{sec:nets}
we develop the syntax and semantics of our proof-net calculus, and
prove  strong normalisation.  We sketch the proofs of faithfulness and full
completeness. More detailed proofs are given in the Appendix.

\paragraph{\textbf{Previous work}} Shirahata has studied a sequent calculus for compact closed categories in \cite{shirahata:1996:comclseq}, while Soloviev has studied natural transformations of definable functors on  compact closed categories with biproducts in \cite{Soloviev:superpositions:1987}.
In \cite{Calco}, the first author has given a comprehensive survey of free constructions for various forms of monoidal category, including traced, compact closed and strongly compact closed categories. Neither biproducts, nor an explicit logical syntax of proof-nets, were considered in that paper.
\section{An Example: Entanglement Swapping}
\label{sec:an-example}

Before proceeding to the details of the proof-calculus and its
categorical model, we  present informally a simple example of a quantum
protocol represented as a proof-net.  First proposed in
\cite{Zukowski1993Event-Ready-Det}, \emph{entanglement swapping} allows two
parties, Alice and Bob, to share an entangled state without directly
interacting with each other.  Instead they each share a Bell pair with
an intermediary, Charlie, who performs a projective Bell-basis
measurement on his part of the two entangled states.  After this
measurement, the qubits retained by Alice and Bob are jointly in a Bell
state, and the outcome of Charlie's measurement will indicate which state it is.

For any finite dimensional vector spaces, $A$, and $B$ there is an
isomorphism  between $A\otimes B$ and the linear maps $A\to B$ via
\[
\sum_{ij}\lambda_{ij} a_i\otimes b_j \isomorphism  a_i \mapsto
\sum_{ij} \lambda_{ij} b_j.
\]
We label states by the maps which they are related to under this
isomorphism.  For example the 4 elements of the Bell basis are
related to the Pauli maps, up to a global phase.  Since
$\frac{1}{\sqrt{2}}(\ket{00}+\ket{11}) \leftrightarrow \id{Q}$, we
represent a Bell state as a proof-net as shown below. 
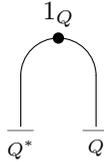
\[
\includegraphics{proofnets_11.ps}
\]
We view this as representing a 2-qubit state, but if read from top to
bottom, it can also been seen as the quantum process which produces
the state.  In general proof-nets are understood as processes but if, as
in this case, the proof-net is \emph{normal} then there is no danger
in identifying the quantum state with the process which prepares it.  

The other component of this protocol is the measurement in the Bell basis.
Suppose that the measurement performed by Charlie yields the state
$X$; then the effect of this measurement is to project his two qubits
onto that state.  This is dual to preparing the state, and represented
by the following diagram fragment
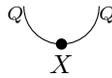
\[
\includegraphics{proofnets_17.ps}
\]
Combining two Bell states and the projection we get the proof-net labelled (a) below
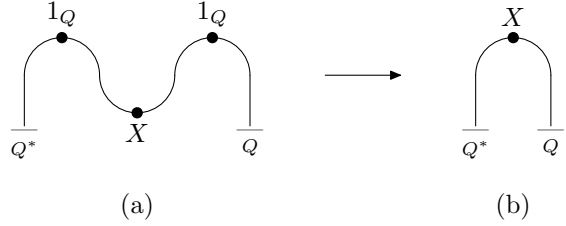
\[
\includegraphics{proofnets_12.ps}
\]
which represents the whole protocol: the preparation of two Bell
states in parallel, and the projection of two of the qubits onto the
state $X$.  This is a process which prepares a 2-qubit state --- we view
measurements as destructive --- and by \emph{normalising} the
proof-net we can compute which state is prepared.  For a proof-net as simple as this one, the result is simply the state which codes the 
composition of the functions labelling the arcs.  In this case the
resulting state is that coded by $\id{Q}\circ X\circ \id{Q} = X$, labelled (b) above.

In reality, there four possible outcomes of a Bell measurement;
these distinct possibilities are represented as \emph{slices}, which
are in a sense different pages of the diagram.  In this case each
slice represents a different element of the Bell basis.
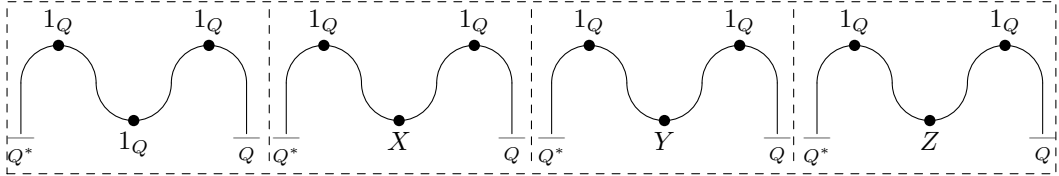
\[
\includegraphics{proofnets_14.ps}
\]
Further, the party who performs the measurement knows which outcome
actually occurred;  we represent this classical information with a
``gearstick''.  In each slice we use a different index to label each
possible outcome.  The final protocol is shown below.
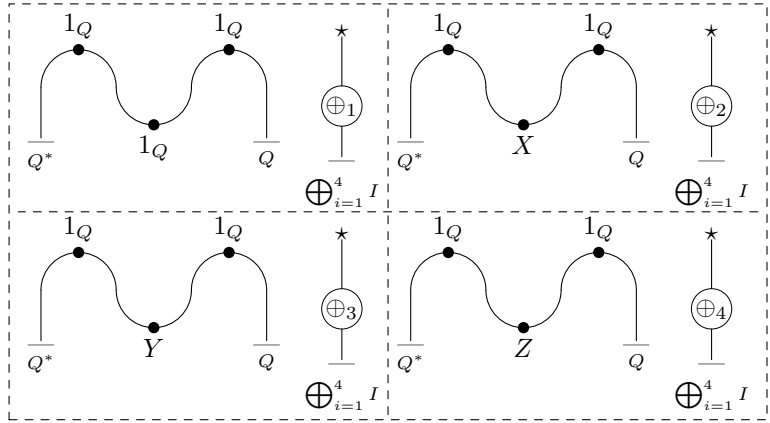
\[
\includegraphics{proofnets_15.ps}
\]
In the following sections we formalise the syntax and semantics of this
proof calculus, and prove that the diagrammatic reasoning employed is
correct with respect to any suitable category.

\section{Categorical Preliminaries}\label{sec:categ-prel}

We recall the definitions and key properties of strongly compact
closed categories with biproducts (SCCCBs).  Considered separately, compact
closure and biproducts are standard structures, and may be found in
\cite{Mitchell:ThCat:1965,MacLane:CatsWM:1971,KelLap:comcl:1980} for
example.  Compact closed categories with biproducts have also been
studied by Soloviev \cite{Soloviev:superpositions:1987} and, with some
strong additional assumptions, as \emph{Tannakian
categories} \cite{Deligne:tannakiennes:1991}.  They have also been studied in a Computer Science context in the first author's work on Interaction Categories \cite{AGN:1996:IntCats}. Strong compact closure
is introduced, and an axiomatic approach to quantum mechanics based on
strongly compact closed categories with biproducts is developed, in
\cite{AbrCoe:CatSemQuant:2004}.  

We will use \fdhilb, the category of
finite dimensional complex Hilbert spaces and linear maps as a running
example.   Another example of an SCCCB is \rel, the category of sets and
relations.


\begin{definition}[Symmetric Monoidal Category]
A \emph{symmetric monoidal} category is a category \catC equipped with a bifunctor 
\[
        - \otimes - : \catC \times \catC \rTo \catC,
\]
a monoidal unit object $I$ and certain natural isomorphisms 
\[
\lambda_A : A \simeq {\rm I}\otimes A\quad\quad\quad\quad\quad\ \ \rho_A: A \simeq
A\otimes{\rm I}
\]
\[
\alpha_{A,B,C}:A\otimes(B\otimes C)\simeq (A\otimes B)\otimes C\vspace{0.5mm} 
\] 
\[
\sigma_{A,B}:A\otimes B\simeq B\otimes A
\]
which satisfy certain coherence conditions \cite{MacLane:CatsWM:1971}.
Without essential loss of generality,  we can assume that $\lambda,
\rho$ and $\alpha$ are all identities; that
is, we can assume a \emph{strict} monoidal category. 
\end{definition}

In any symmetric monoidal category $\catC$, the
endomorphisms $\catC (I , I )$ form a commutative monoid \cite{KelLap:comcl:1980}.
We call these endomorphisms the \emph{scalars} of \catC.   For each
scalar $s:I\to I$ we can define a natural transformation   
\begin{diagram} 
s_A :A =  I \otimes \!A & \rTo^{s \otimes 1_A} & I
\otimes\! A =  A\,.
\end{diagram}
Hence, we can define \emph{scalar multiplication} $s \bullet f := f \circ s_A=s_B\circ f$ for $f :
A \rightarrow B$.  Then we have 
\[
(s \bullet g)\circ(r \bullet f)=(s\circ r)\bullet(g\circ f)
\]
for $r:I\to I$ and $g:B\to C$. 

\begin{definition}[Compact Closed Category]A symmetric monoidal category is \emph{compact closed} if to each object $A$ there
is an assigned left adjoint $(A^*, \eta_A, \epsilon_A)$ such that the composites
\[
\begin{array}{c}
A = A\otimes I \rTo^{\id{\!A}\otimes\eta_A} A\otimes A^*\otimes A
 \rTo^{\epsilon_A\otimes\id{\!A}} I\otimes A
= A \\
A^* = I \otimes A^* \rTo^{\eta_A\otimes\id{\!A^*}} A^*\!\otimes
A \otimes A^* 
\rTo^{\id{\!A^*}\!\otimes\epsilon_A}  A^* \! \otimes I = A^*
\end{array}
\]
are both identities.
\end{definition}

In \fdhilb the tensor product is just the usual Kronecker tensor
product, and $I = \mathbb{C}$.    Since any linear map from $\mathbb
C$ to itself is fixed by its value at 1, the formal scalars in \fdhilb
are indeed the complex numbers.  If $A$ is some finite dimensional
Hilbert space then, we can take $A^*$ to be the usual dual, the space
of linear maps $A\to \mathbb C$. Given a basis $\{a_i\}_i$ for $A$, and its dual basis $\{\overline{a_i}\}_i$, the required maps are
  \begin{eqnarray*}
    \eta_A : 1 &\mapsto& \sum_i \overline{a_i}\otimes a_i;\\
    \epsilon_A : a_i \otimes \overline{a_j} &\mapsto & \delta_{ij}.
  \end{eqnarray*}  
A routine calculation verifies that the required equalities hold,
\fdhilb is indeed compact closed.

For each morphism $f:A\to B$ in a compact closed category we can construct its
\emph{name}, $\name{f}:I\to A^*\otimes B$, \emph{coname},
$\coname{f}:A\otimes B^* \to I$, and \emph{dual}, $f^*:B^*\to A^*$, by
\[
\begin{diagram}
I & \rTo^{\eta_A} & A^*\otimes A \\
& \rdTo_{\name{f}} & \dTo_{\id{A^*} \otimes f} \\
&& A^*\otimes B
\end{diagram}
\qquad\qquad\qquad
\begin{diagram}
A\otimes B^*  && \\
\dTo^{f \otimes \id{B^*}} & \rdTo^{\coname{f}} & \\ 
B\otimes B^* & \rTo_{\epsilon_A} & I
\end{diagram}
\]
\begin{diagram}
B^* & = & I \otimes B^*& \rTo^{\quad\eta_A\otimes\id{B^*}\quad} & A^*\otimes A \otimes B^* \\
\dTo^{f^*} &&&& \dTo_{\id{A^*}\otimes f\otimes\id{B^*}} \\
A^* & = &  A^* \otimes I & \lTo^{\quad\id{A^*}\otimes\epsilon_B\quad} & A^*\otimes B \otimes B^*
\end{diagram}
In particular, the map $f \mapsto f^*$ extends to a contravariant
endofunctor with $A \iso A^{**}$.

Each compact closed category admits a categorical trace. That is, for every
morphism $f:A\otimes C\to B\otimes C$ certain axioms
\cite{JSV:TraMonCat:1996} are satisfied by $\text{Tr}_{A,B}^C(f):A\to
B$, defined as the composite: 
\begin{diagram}
A = A\otimes I & \rTo^{\id{A}\otimes \eta_{C^*}} & 
A \otimes C \otimes C^* & \rTo^{f\otimes \id{C^*}} & 
B \otimes C \otimes C^* & \rTo^{\id{B}\otimes\epsilon_C} &
B \otimes I = B.
\end{diagram}

\noindent The following results are proved in \cite{AbrCoe:CatSemQuant:2004}.

\begin{lemma}\label{lem:fourlems}
Suppose we have maps $E\rTo^k A\rTo^f B \rTo^g C \rTo^h D$.  Then we
have the following equations.
\renewcommand{\theenumi}{(\alph{enumi})}
\begin{enumerate}
\item Absorption:\label{lem:absorb}
\[
(\id{A^*}\!\!\otimes g)\circ \name{f} =\name{ g\circ f}
\]
\item Backward absorption:\label{lem:bkabsorb}
\[
(k^*\otimes \id{A^*}\!)\circ \name{f}=\name{f\circ k}
\]
\item Compositionality:\label{lm:compos}
\[
\lambda^{-1}_C\circ (\coname{f}\otimes
\id{C})\circ(\id{A}\otimes\name{g})\circ\rho_A=g\circ f
\]
\item Compositional cut:\label{lem:compCUT}
\[
(\rho^{-1}_A\!\otimes \id{D^*}\!)\circ(\id{A^*}\!\otimes\coname{g}\otimes\!
\id{D})\circ({\name{f}}\otimes{\name{h}})\circ\rho_I =\name{h\circ
g\circ f}
\]
\end{enumerate}
\end{lemma}
\noindent The obvious analogues of Lemma~\ref{lem:fourlems}\ref{lem:absorb} and  \ref{lem:fourlems}\ref{lem:bkabsorb} for conames also hold.

\begin{definition}[Zero Object]
A \emph{zero object} in \catC is both initial and terminal.  If $\ZERO$ is
a zero object, there is an  arrow $0_{A,B}:A \rTo \ZERO  \rTo  B$ 
between any pair of objects $A$ and $B$.
\end{definition}

\begin{definition}[Biproduct]
Let $\catC$ be a category with a zero object and binary products and coproducts.
Any arrow 
\[
A_1 \coprod A_2 \rightarrow A_1 \prod A_{2}
\]
can be written uniquely
as a matrix $(f_{ij})$, where $f_{ij} : A_{i} \rightarrow A_j$.
If the arrow 
\[ \left( \begin{array}{cc}
1 & 0 \\
0 & 1
\end{array} \right) \]
is an isomorphism for all $A_1$, $A_2$, then we say that $\catC$ has
\emph{biproducts}, and write $A \oplus B$ for the biproduct of $A$ and 
$B$. 
\end{definition}

If $\catC$ has biproducts then we can define an operation of addition
on each hom-set $\catC (A, B)$ by 
\[ 
f+g = \nabla \circ ( f\oplus g) \circ \Delta
\]
for  $f,g:A\to B$, where $\Delta=\langle 1_A,1_A\rangle$  and
$\nabla=[1_B,1_B]$. This operation is associative and commutative, with $0_{AB}$ as a unit. Moreover, composition is bilinear with respect to this
semi-additive structure.

In \fdhilb the direct sum of Hilbert spaces gives a biproduct, with
the vector space $\{0\}$  as the zero object; the addition
on hom sets is normal addition of maps. In \rel the biproduct is given by disjoint union of sets; addition of relations is given by their union.

If $\catC$ has biproducts, we can choose projections $p_1$, $p_2$ and
injections $q_1$, $q_2$ for each $A \oplus B$ satisfying:
\[ p_i \circ q_j = \delta_{ij} \qquad q_1 \circ p_1 + q_2 \circ p_2 =
1_{A \oplus B} \] 
where $\delta_{ii} = \id{}$, and $\delta_{ij} = 0$, $i \neq j$.

\begin{remark}We note that if \catC is already equipped with a
semiadditive structure as above then one can define the biproduct
directly as a diagram,
\begin{diagram}
A & \pile{\lTo^{p_1}\\\rTo_{q_1}} & A\oplus B & \pile{\rTo^{p_2}\\\lTo_{q_2}}&B
\end{diagram}
satisfying
\[ 
p_i \circ q_j = \delta_{ij} 
\qquad 
q_1 \circ p_1 + q_2 \circ p_2 = \id{A \oplus B}.
\] 
This fact will be used for the biproduct structure of the proof calculus.
\end{remark}

Of course, the biproduct defines a monoidal structure, with unit
object $\ZERO$.  As before, we will take it to be strict:
\[
(A\oplus B)\oplus C = A\oplus (B\oplus C), \qquad\qquad \ZERO\oplus A = A = A \oplus\ZERO.
\]

\begin{proposition}[Distributivity of $\otimes$ over $\oplus$]\label{distributivity}
In monoidal closed categories with biproducts there are natural isomorphisms 
\[
d_{A,B,C}:A\otimes(B\oplus C)\iso (A\otimes B)\oplus(A\otimes C)
\]
\[
d_{A,\cdot,\cdot}\!\!=\langle 1_A\otimes p_1,1_A\otimes p_2\rangle\quad\ \ 
d_{A,\cdot,\cdot}^{-1}\!\!=[ 1_A\otimes q_1,1_A\otimes q_2]\,.
\]
A left distributivity isomorphism can be defined similarly.
\end{proposition}

\begin{proposition}\label{prop:zero-monoid}
  In a monoidal closed category with a $\ZERO$ object there are natural isomorphism $A\otimes\ZERO \isomorphism \ZERO \isomorphism \ZERO \otimes A$.
\end{proposition}

\begin{proposition}[Self-duality for $(\cdot)^*$]\label{prop:selfdual}
In a compact closed category with biproducts the following natural isomorphisms exist.
\[
\begin{array}{ccccc}
A^{**} \iso A & \qquad & (A\otimes B)^* \iso A^*\otimes B^* & \qquad & I^* \iso I \\
&& (A\oplus B)^* \iso A^* \oplus B^* & \qquad & \ZERO^* \iso \ZERO
\end{array}
\]
\end{proposition}
\noindent It will be notationally convenient to take all the canonical
maps of Prop.~\ref{prop:zero-monoid} and \ref{prop:selfdual} as equalities.

\begin{definition}[Strong Compact Closure]
A compact closed category \catC is \emph{strongly compact closed} if 
the assignment $A \mapsto A^*$ extends to a \emph{covariant} involutive
compact closed functor.  Write $f_*$ for the action of this functor on
arrow $f$. (See \cite{Abramsky:2005:apt} for an alternative definition
of strong compact closure).
\end{definition}

Given $f:A\to B$ in a strongly compact closed category \catC we can define its
\emph{adjoint} $f\dagg : B \to A$ by $f\dagg = (f_*)^* = (f^*)_*$.
The assignments $A\mapsto A$ on objects and $f\mapsto f^\dag$ on
arrows define an involutive functor .  If \catC has biproducts then
$(\cdot)^\dag$ preserves them, and hence is additive.

If \catC is strongly compact closed and has biproducts we require a
compatibility condition, namely that the coproduct injections 
\[
q_i: A_i\to \bigoplus_{k=1}^{k=n}A_k
\]
satisfy $q_j^\dagger\circ q_i=\delta_{ij}$. It then follows that the
projections and injections additionally satisfy $(p_i)^{\dagger}=q_i$.

In \fdhilb the $(\cdot)^\dag$ is the usual adjoint of a linear map,
given by 
\[
\innp{\psi}{f\phi} = \innp{f^\dag\psi}{\phi}.
\]
The functorial action $f_{*}$ is defined by
\[ f_{*}(\phi)(v) =\phi \circ f^{\dag}(v) . \]
\begin{remark}
  In \cite{AbrCoe:CatSemQuant:2004},  $A^*$ is defined to be the
  \emph{conjugate space} of $A$, which has the advantage of being strictly involutive. 
\end{remark}


\section{The Logic of SCCCBs}\label{sec:logic-scccbs}

The purpose of this paper is to present a logic whose syntax captures
the structure of the free strongly compact closed category with biproducts generated by a category with an involution.  The formulae of the logic
represent the objects of the free category, while the proofs represent the
arrows\footnote{See the remarks at the end of
\cite{KelLap:comcl:1980} which explain why a description of the free compact closed category is the strongest available form of coherence theorem  for such categories.}.

Let $F$ be the functor which takes a category with involution to the free strongly
compact closed category with biproducts generated upon it.
\begin{diagram}
\textbf{InvCat} & \pile{ \rTo^{\qquad F \qquad} \\ \bot \\ \lTo_U} & \textbf{SCCCB}
\end{diagram}
Here $\InvCat$ is the category of categories with involutions, \ie identity-on-objects, contravariant, involutive functors, and functors preserving the given involutions.
We will define a logic relative to a ground category
\catA (standing for ``axioms'' or ``atoms'' according to taste) with involution $(\cdot)^{\dag}$.
The objects of \catA will form the atomic formulas of the 
syntax, and its arrows will give non-logical axioms.  The formulas of the
resulting logic will represent the objects of the generated category 
$F\catA$, while the proofs will represent its arrows.

We can simplify our task, following \cite{Calco}.  It is shown there that $F_{\mathrm{CC}}$, the functor that constructs the free compact closed category generated by a category, lifts to $\InvCat$ to yield the free strongly compact closed category over a category with involution. This amounts to the observation that, given an involution on the base category, it lifts to one on the freely generated compact closed category, and moreover this lifted involution is compatible with the compact closed structure in the required fashion---so that, in  particular, 
\[ \epsilon_{A} = \sigma_{A^{*},A} \circ \eta_{A}^{\dag} . \]
Lemma \ref{scclemma} in the Appendix recalls how the involution is lifted.
  
Henceforth we will assume that \catA has an adjoint $f^\dag$ assigned to every
arrow $f$.

\begin{definition}
The \emph{formulae} of the logic are built from the following grammar:
\[ 
F ::= \ZERO \; | \; \I \; | \; A \; | \; A^* \; | \; F\otimes F \; | \;F \oplus F,
\]
where $A$ ranges over the objects of \catA, which we shall refer to as
\emph{atoms}.  In order to capture the strictness of the connectives with respect to their units,
the use of the units is restricted:  $\ZERO$ may not occur as a
subformula of any formula other than itself;  while $\I$ may only
occur immediately under a biproduct, i.e. $\I \otimes A$ is banned, but
$(\I\oplus A)\otimes B$ is permitted.
While it is technically convenient to admit $\I$ as a valid formula, 
a correctness condition for proof-nets will guarantee that $\I$ never
occurs in a conclusion of a correct proof without an accompanying $\oplus$. 
We define $(\cdot)^*$ on arbitrary formulae by  the following equations:
\[
\begin{array}{rcl}
X^{**} & =  & X \\
(X\otimes Y)^*  &  =  & X^* \otimes Y^* \\
(X \oplus Y)^*  &  =  & X^* \oplus Y^*\\
\I^* & = & \I\\
\ZERO^* & = & \ZERO.
\end{array}
\]
\end{definition}

We use the notation convention that upper case letters $A,B,C$ from the
start of the latin alphabet are atoms and those from the end of
the alphabet $X,Y,Z$ are arbitrary formulae.  Upper case Greek
letters $\Gamma,\Delta,\Sigma$ signify lists of formulae.

\begin{definition}
We shall use \emph{axiom} synonymously with  \emph{arrow of \catA}.
\end{definition}

Cyclic structures play an important role in the theory of compact closed categories
and we shall have need of them in the syntax.
Define the set of \emph{endomorphisms} $E(\catA)$ by the disjoint
union 
\[
E(\catA) = \sum_{A \in |\catA|} \catA(A,A),
\]
and let the set of \emph{loops} $[\catA]$ be the quotient of
$E(\catA)$ generated by the relation $f\circ g \sim g\circ f$ whenever
$A\rTo^f B \rTo^g A$.


\section{Proof-nets}\label{sec:nets}

We present a graphical proof notation which captures
precisely the structure of strongly compact closed categories with biproducts; in fact we offer a
faithful and fully complete representation of $F\catA$.

\subsection{Syntax}

\begin{definition}[slice]
A \emph{slice} is a finite oriented graph with edges labelled by
formulae.  The graph is constructed by composing the following
nodes, which we call \emph{links}, while respecting the labelling on the
incoming and outgoing edges. 
\begin{description}
\item[Axiom : ] No incoming edges; two out-going edges.    The link itself is
labelled by an axiom  $f:A\to B$ .  One outgoing edge is labelled
$A^*$, the other, $B$. 
\item[Cut : ] Two incoming edges; no outgoing edges.  Each cut is
labelled either by an axiom $f:A\to B$ with incoming edges are labelled 
by atoms $A$ and $B^*$, or else it is labelled by an identity with the
incoming edges labelled by $X$ and $X^*$ for an arbitrary formula $X$.
\item[Times : ] Two incoming edges labelled $A$ and $B$; one outgoing
edge labelled $A\otimes B$.
\item[Plus 1 : ] One incoming edge labelled $A$; one outgoing edge
labelled $A\oplus B$.
\item[Plus 2 : ] One incoming edge labelled $B$; one outgoing edge
labelled $A\oplus B$.
\item[I : ] No incoming edges; one outgoing edge labelled by $\I$.
\end{description}
The orientation is such that edges enter the node from the top, and
exit from the bottom.  The \emph{conclusions} of the slice are
those labels on outgoing edges of links which are left unconnected.
The order of the conclusions is significant.
There is one correctness criterion:  every $\I$-link must be connected
to either a Plus-link or a cut labelled with $\id{\I}$.
\end{definition}

\begin{definition}[net]
A \emph{net} (or proof-net) is a finite multiset of slices where
each slice has the same conclusions.  The conclusions of the net are
the same as those of its slices.
\end{definition}

We emphasise that empty slice is a valid slice, having no conclusions,
and the empty set of slices is a valid net.  In particular the empty net may be
considered as having any conclusions;  since there is no rule for
introducing it otherwise, the additive unit $\ZERO$ can only occur
among the conclusions of an empty net.

\begin{example}
This net represents the distribution of $\otimes$ over $\oplus$.
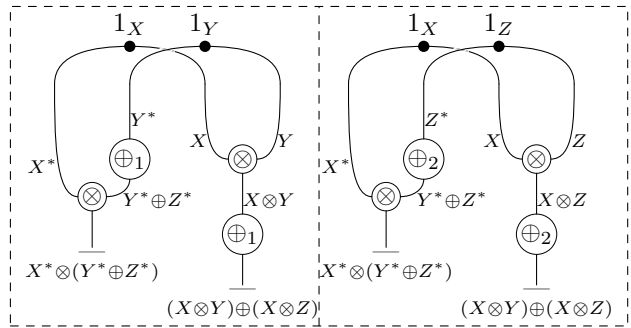
\[
\includegraphics{proofnets_16.ps}
\]
\end{example}

\begin{definition}[Normal Forms]\label{defn:normalform}
A slice is \emph{normal} if every connected component either has no
cut links, or is a closed loop formed by an axiom link and an identity
cut. We identify loops if their labels are related by the equivalence relation on endomorphisms give in section ~\ref{sec:logic-scccbs}.  A net is normal if every slice is normal.
\end{definition}

\begin{definition}[$\beta$-Reduction]\label{defn:betareduction}
Let $\to_\beta$ be the reflexive transitive closure of the relation
defined on slices by the following set of rewrites on cut links.
\begin{enumerate}
\item A cut between atomic formulae.  Atomic formulae are only
introduced by axiom links, so there are two subcases.
\begin{enumerate}
\item If both formulae belong to the same axiom (say $f$):
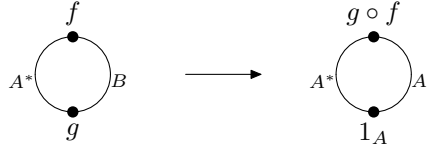
\[
\includegraphics{proofnets_3.ps}
\]
\item If the cut formulae are conclusions of different axioms, say
$f$ and $h$:
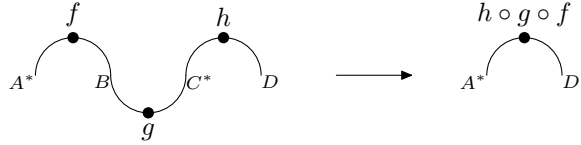
\[
\includegraphics{proofnets_4.ps} 
\]
\end{enumerate}
\item Cut between two tensor products: 
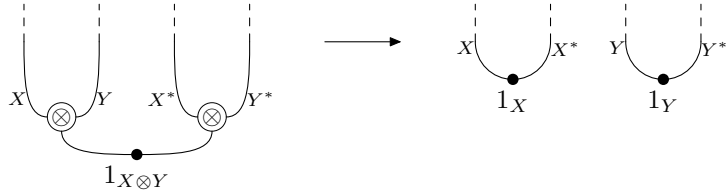
\[
\includegraphics{proofnets_5.ps}
\]
\item Cut between two biproducts:
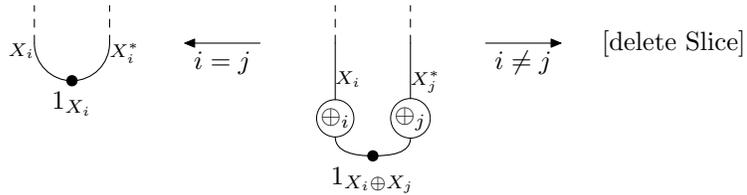
\[
\includegraphics{proofnets_6.ps}
\]
\item Cut between two $\I$-links:
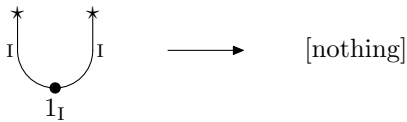
\[
\includegraphics{proofnets_10.ps}
\]
\end{enumerate}
Extend $\to_\beta$ to proof-nets by $\pi\to_\beta\pi'$ iff there is an
injective map $p$ from the slices of $\pi'$ to those of $\pi$, such
that if $p(s') = s$ then $s\to_\beta s'$, and for every slice $s$ of
$\pi$ not in the image of $p$, there is a $\to_\beta$ sequence ending
in ``Delete Slice''.  Let $=_\beta$ be the symmetric closure of
$\to_\beta$.
\end{definition}

\begin{theorem}[Cut Elimination]\label{thm:netsCE}
The relation $\to_\beta$ is confluent and terminating;  further the
$\beta$-normal forms are normal in the sense of
Def.~\ref{defn:normalform} above.
\end{theorem}

\begin{figure} 
\topfigrule
\figureline
\begin{center}
\includegraphics{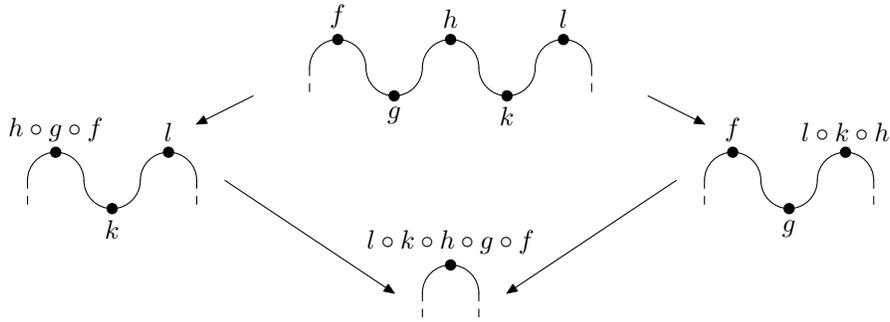}
\end{center}
\caption{Confluence of cut elimination step 1(b)}\label{fig:confluence}
\figureend
\end{figure}

\begin{proof}
It suffices to consider $\to_\beta$ on slices alone.

In the case of 1(b) pairs of rewrites may
interfere as shown in figure \ref{fig:confluence}.  However
associativity in the underlying category \catA prevents any
conflict. 

Case 1(a) may also conflict with 1(b) as shown in figure \ref{fig:confluenceii}; in this case the two resulting components are identified by the equivalence on loops.

With these exceptions, each reduction step is purely
local --- no rewrite can affect any other --- hence the process is
confluent.    Since each step reduces the complexity of the net, there
is no infinite reduction sequence, and hence every net is strongly
normalising.

The only situation where a cut will not be eliminated are those in case
1(a); hence when no more rewrites can be done the slice is normal, as
required.
\end{proof} 

\begin{figure} 
\topfigrule
\figureline
\begin{center}
\includegraphics{proofnets_19.ps}
\end{center}
\caption{Confluence of cut elimination step 1(a)/1(b)}\label{fig:confluenceii}
\figureend
\end{figure}

\subsection{Semantics}

\begin{definition}[Semantics of proof-nets]\label{def:netsem}
Let $\nu$ be a proof-net with conclusions $\Gamma$.  Define
an arrow of $F\catA$, $\denote{\nu}:I\to\bigotimes\Gamma$, by
recursion on the structure of $\nu$.  Consider each slice $s$ of $\nu$.
\begin{itemize}
\item If $s$ is just an axiom link corresponding to the arrow $f:A\to
B$, then let $\denote{s} = \name{f}:I\to A^*\otimes B$.
\item If $s$ has several disconnected components
$s_1,\ldots,s_n$ then define 
\[
\denote{s} =  \bigotimes^n_{i=1}\denote{s_i}.
\]
\item If $s$ is built by applying a cut labelled by $f:A\to B$ 
between conclusions $A$ and $B^*$ of $s\ds$, suppose that we have
constructed $\denote{s\ds}:I\to \Gamma\otimes A \otimes B^*\otimes
\Delta$. Then define $\denote{s}$ by the composition 
\begin{diagram}
I & 
\rTo^{\quad\denote{s\ds}\qquad} & 
\Gamma\otimes A \otimes B^*\otimes \Delta &
\rTo^{\quad\id{\Gamma}\otimes\coname{g}\otimes\id{\Delta}\quad} &
\Gamma\otimes\Delta.
\end{diagram}
\item If $s$ is built by applying a $\otimes$-link  between conclusions $A$
and $B$ of $s\ds$ then let $\denote{s} = \denote{s\ds}$.
\item If $s$ is built by applying a $\oplus_i$ link to conclusion $A_j$
 of $s\ds$, construct $\denote{s\ds}:I\to \Gamma\otimes A_j \otimes
\Delta$ then define $\denote{s}$ by the composition
\begin{diagram}
I & 
\rTo^{\quad\denote{s\ds}\qquad} & 
\Gamma\otimes A_j \otimes \Delta &
\rTo^{\quad\id{\Gamma}\otimes q_i\otimes\id{\Delta}\quad} &
\Gamma\otimes(A_1\oplus A_2)\otimes\Delta.
\end{diagram}
\item If $s$ is an $\I$-link, then $\denote{s} = \id{I}$.
\item If $s$ is the empty slice $\denote{s} = \id{I}$.
\end{itemize}
All these constructions commute wherever the required compositions
are defined due to the functoriality of the tensor, hence 
$\denote{s}$ is well defined.
Let $\nu$ be the net composed of the slices $s_1,\ldots,s_n$.  We
define
\[
\denote{\nu} = \sum_{i=1}^n \denote{s_i}.
\]  
If $\nu$ is the empty proof-net (i.e. it has no slices) $\denote{\nu}
= 0_{I,\Gamma}$.
\end{definition}

\begin{theorem}[Soundness]
If a net $\nu \to_\beta \nu'$ then $\denote{\nu} = \denote{\nu'}$.
\begin{proof}
Each of the one step rewrite rules preserves
denotation.  For each rewrite rule we show the corresponding equation.
\begin{enumerate}
\item Suppose we have arrows $B \rTo^e A \rTo^f B \rTo^g C \rTo^h D$
in \catA; then
\begin{enumerate}
\item We have 
\[
\begin{array}{rcl}
\coname{e} \circ \name{f} 
        & \quad = \quad & \epsilon_{A^*} \circ ( \id{A^*}\otimes e) \circ (\id{A^*} \otimes f ) \circ \eta_A \\ 
        & = & \epsilon_{A^*} \circ (\id{A^*} \otimes (e\circ f ) \circ\eta_A  \\
        & = & \epsilon_{A^*} \circ \name{e\circ f}
\end{array}
\]
directly from the definition of the name and coname.
\item The required equation 
\[
        ( \id{A^*}\otimes\coname{g}\otimes\id{D}) \circ (\name{f}\otimes\name{h}) 
        = 
        \name{h\circ g\circ f}
\]
is lemma \ref{lem:fourlems}.\ref{lem:compCUT} verbatim.
\end{enumerate}
\item The case for tensor follows from $\epsilon_{A\otimes B} =
\sigma\circ (\epsilon_A\otimes\epsilon_B)$.
\item By using forwards and backwards absorption (lemmas
\ref{lem:fourlems}.\ref{lem:absorb},\ref{lem:fourlems}.\ref{lem:bkabsorb}) we have
\[
        \epsilon_{A_i\oplus A_j} \circ (q_i \otimes q_j) 
        = \coname{p_j \circ \id{A_i\oplus A_j} \circ q_i} 
        = \left\{ \begin{array}{ll} 
                        \epsilon_{A_i} & \text{ if } i=j \\ 
                        \coname{0_{A_i,A_j}} & \text{ if } i \neq j \\
          \end{array}\right.
\]
Consider the case where $i\neq j$.  We note that $\coname{0_{A_i,A_j}}
= 0_{A_i\otimes A_j^*,I}$, and since any arrow composed with, or tensored
with, a zero map is itself a zero map the denotation of the entire
slice must be zero.  Hence we may delete it without altering the
denotation of the net.
\item Since $F\catA$ is strict, we have that $\epsilon_I \circ (\id{I}\otimes\id{I}) = \id{I}$ as required.
\end{enumerate}
The result follows by the functoriality of the tensor.
\end{proof}
\end{theorem}

Now we  show that the proof-net syntax is a faithful representation of
the category $F\catA$.  For the purposes of the following proof,  by
\emph{involution on a set $X$} we will mean a category consisting of a finite
coproduct of copies of the category \textbf{2} (with objects $0$, $1$,
and one non-identity arrow $0 \rightarrow 1$), whose objects are in
bijective correspondence with the elements of $X$.

\begin{lemma}
In order to specify a normal slice uniquely the following data are required:
\begin{enumerate}
\item The list of conclusions $\Gamma$;
\item A list of booleans $B$, indicating, for each occurrence of the
connective $\oplus$ in $\Gamma$, whether the left or right subformula
was the premise of the link which introduced it;
\item An involution $\theta$ on those atoms of $\Gamma$ which are not
  introduced by the $\oplus$ rules such that each atom is paired with
  a copy of $0$ iff it is negative, together with a  functor $p:\theta
  \to \catA$;
\item A multiset $L$ of loops in \catA.
\end{enumerate}  
\end{lemma}
\begin{proof}
  Clearly every normal slice will define the four data above, and do
  so uniquely.  We show how to reconstruct the the slice from the data.
  For each formula $X$ of $\Gamma$, the syntax of $X$ combined with
  $B$ fix a unique set  of logical ($\otimes$, $\oplus$, $\I$) links
  which derive the formula from its constituent atoms.  Necessarily,
  there are an equal number of positive and negative atoms; $\theta$
  specifies an arrangement of axiom links between them labelled by
  $p$.  This doesn't totally fix the slice, since we may have
  additional disconnected components.  Since they have no conclusions,
  and the slice is normal, any remaining components must be loops,
  which are specified by $L$.
\end{proof}

\begin{theorem}[Faithfulness]
If nets $\nu, \nu\ds$ have the same conclusions $\Gamma$ then 
$\denote{\nu} = \denote{\nu\ds}$ implies $\nu =_\beta \nu'$.
\begin{proof}
Let $s$ be a normal slice.    By Def.~\ref{def:netsem}
$\denote{s} = c\bullet f$, where $c$ is a scalar and $f$ has the
following structure:
\[
I
\rTo^{\name{f_1}\otimes\cdots\otimes\name{f_n}} 
\bigotimes_{i=1}^n(A_{2i-1}^*\otimes A_{2i}) 
\rTo^\sigma 
\bigotimes_{i=1}^{2n} A_{\sigma(i)} 
\rTo^\kappa \Gamma
\]
upto a scalar multiple, where $\sigma$ is a permutation, and $\kappa$
is a tensor product of 
identities and injections.  This structure suffices to define the
data of the preceding lemma.  

Every $\oplus$ in $\Gamma$ must be introduced by the injections
$\kappa$, which serve to define $B$.  Given $\sigma$ and any ordering on the names
$\name{f_i}$ the pair $(\theta, p)$ is easily reconstructed.  

If $c \neq \id{I}$ then the free construction of $F\catA$ guarantees
that it is product of arrows of the form
\[
        I \rTo^{\name{l}} A^*\otimes A \rTo^{\epsilon_{A^*}} I.
\]
Each automorphism $l$ defines a loop, so $c$ defines $L$.
It is easy to verify that the slice reconstructed from
$c$ and $f$ will be the original slice $s$. 

Since the addition is freely constructed,  $\denote{\nu} =
\denote{\nu'}$ implies that both are equal to the same formal sum
$\sum_i f_i$, where each $f_i$ is the denotation of a slice.  Since
each $f_i$ determines a unique normal slice, we have that the normal
forms of $\nu$ and $\nu'$ comprise the same multiset of slices, hence
$\nu =_\beta \nu'$.
\end{proof}
\end{theorem}

It should be noted that the faithfulness result required the
conclusions of the nets to be specified.  In fact the syntax is not
truly injective onto the arrows of $F\catA$.  For example, the
proof-nets
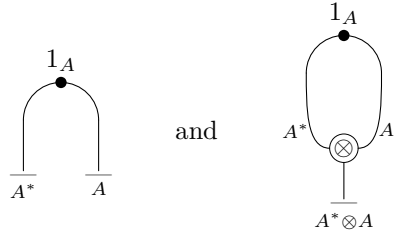
\[
\raisebox{4mm}{\includegraphics{proofnets_8.ps}}
\raisebox{12mm}{\qquad\text{ and }\qquad}
\includegraphics{proofnets_9.ps}
\]
both denote the map $\eta_A$.

Let $F_{\mathrm{SCC}}: \textbf{Cat}\to\textbf{ComClCat}$ be the functor which takes
a category to the free compact closed category generated by it.  This functor
has been described in detail in \cite{KelLap:comcl:1980}.  We
note that $F_{\mathrm{SCC}}\catA$ is a subcategory of $F\catA$.

\begin{theorem}[Full Completeness]
Let $f$ be an arrow of $F\catA$, the free compact closed category with biproducts on \catA;  then
there exists a net $\nu$ such that $\name{f} = \denote{\nu}$.
\begin{proof}[Sketch proof]
We note that each object of $F\catA$ is canonically isomorphic to an
object in additive normal form; that is where no occurrence of $\oplus$
occurs in the scope of any occurrence of $\otimes$.  Hence given an
arrow $f:A\to B$ in $F\catA$ one can construct the three other sides of
following square 
\begin{diagram}
A & \rTo^f & B \\
\dTo^\iso && \uTo_\iso\\
\bigoplus_i \bigotimes_{j_i} A_{j_i} & \rTo_{\qquad\bigoplus_i f_i\qquad} & \bigoplus_{i} \bigotimes_{k_i} B_{k_i}
\end{diagram}
such that each $f_i$ is a (possibly empty) sum of arrows of
$F_{\mathrm{SCC}}\catA$.  The theorem of Kelly-Laplaza 
\cite{KelLap:comcl:1980} gives an explicit 
description the arrows in $F_{\mathrm{SCC}}\catA$ and hence immediately a proof-net
for each one. Then each $f_i$ yields a collection of slices.  From
here it is straightforward to construct the proof-net of $\bigoplus_i
f_i$.
\end{proof}
\end{theorem}

The appendix contains a more detailed proof of the theorem.


\section{Further Work}

In the present paper, we have focused on freely generating the
structure over a category with no additional structure.  If the
category in question has an object for the type of qubits, then the
resulting free structure will not contain, for example, the controlled
not gate, nor any other multi-qubit operation.  Without such maps the
expressivity of the system is limited.  Therefore an important further
step is to consider the structure of the freely generated strongly
compact closed category with biproducts over a category with a
\emph{given} symmetric monoidal structure.  This is carried out in the
forthcoming thesis \cite{Duncan:thesis:2006} of the second author.

Furthermore, although it has not been discussed  here, the model may be
further tuned by the choice of the semiring of scalars $I\to I$.  
These represent the ``amplitudes'' of the different terms of a state, 
and hence give rise to the probabilities of different outcomes of a
quantum process.  The equational structure of the scalars constrains
the representable quantum processes.  For example it is known that the
category \catRel does not have enough scalars to represent the
teleportation protocol.  It is possible to specify a desired semiring $R$ of scalars as a separate parameter to a free construction, together with a map from the loops of $\catA$ to $R$ \cite{Calco}.
Given a suitable
rewriting theory for $R$, this can be combined with the proof-net calculus to yield a system in which it should be possible to extract concrete
probabilities and other quantitative information. These ideas will be developed in future work.

\bibliography{all}


\appendix

\section{Proof of  full completeness}

Firstly, we remark that there is a well-known description of the free construction $B \CC$ of a category with finite biproducts generated by a category $\CC$ (for which see e.g. \cite{MacLane:CatsWM:1971}). The objects of $B \CC$ are finite tuples of objects of $\CC$, written $\bigoplus_{i =1}^{n} A_{i}$; morphisms $\bigoplus_{i =1}^{n} A_{i} \rightarrow \bigoplus_{j =1}^{m} B_{j}$ are $n \times m$ matrices whose components are finite multisets of arrows $A_{i} \rarr B_{j}$, \ie elements of the free Abelian monoid generated by $\CC (A_{i}, B_{j})$. Composition is by ``matrix multiplication'', with the composition of $\CC$ bilinearly extended to multisets. This construction, as is also well-known, extends to the construction of free \emph{distributive biproducts} over monoidal categories, with the tensor defined on $B \CC$ by distributivity:
\[ (\bigoplus_{i =1}^{n} A_{i}) \otimes (\bigoplus_{j =1}^{m} B_{j}) = \bigoplus_{i, j} A_{i} \otimes B_{j} . \]
The following is a straightforward extension of this standard result:
\begin{proposition}
The matrix construction lifts to strongly compact closed categories.
\end{proposition}
\begin{proof}
The adjoint of a matrix $(m_{ij})$ is $(m_{ji}^{\dag})$, where the adjoint of the generating strongly compact closed category is applied pointwise to the multiset $m_{ji}$.
The unit for $\bigoplus_{i =1}^{n} A_{i}$ is the diagonal matrix with diagonal elements $\{ \eta_{A_{i}} \}$.
\end{proof}
This  yields a factorization of the adjunction
\begin{diagram}
\textbf{InvCat} & \pile{ \rTo^{\qquad F \qquad} \\ \bot \\ \lTo_U} & \textbf{SCCCB}
\end{diagram}
as
$F = B \circ F_{\mathrm{SCC}}$:
\begin{diagram}
\textbf{InvCat} & \pile{ \rTo^{\qquad F_{\mathrm{SCC}} \qquad} \\ \bot \\ \lTo_{U_{\mathrm{SCC}}} } & \textbf{SCCC} & \pile{ \rTo^{\qquad B \qquad} \\ \bot \\ \lTo_{U_{\mathrm{SCCB}}} } & \textbf{SCCCB}.
\end{diagram}
This factorization underlies the structure of the following argument.
One particular consequence we shall use is the following:
\begin{proposition}
\label{FCCfaithembedprop}
 $F_{\mathrm{SCC}}(\CC)$ embeds faithfully in $F(\CC)$.
\end{proposition}

We will refer to the objects of $\catA$, their images under
$(\cdot)^*$ and the constants \textbf{0} and $I$ as the
\emph{literals} of $F\catA$.   Since $F\catA$ is freely generated, its
objects are formed from the literals by repeated application of the
functors $(-\otimes -)$ and $(-\oplus -)$.  Hence any object may be
described by such a functor and a vector of literals.
For the rest of the section, it will
be understood that by functor we refer only to those constructed from
tensors and biproducts\footnote{Soloviev has treated the
natural transformations of such functors in detail \cite{Soloviev:superpositions:1987},
but here we are only interested in one particular case.}.
Let $\otimes_n : F\catA \times \cdots \times F\catA \to F\catA$ be the
$n$-fold tensor; similarly let $\oplus_n$ be the $n$-fold biproduct.
Call $N$ a \emph{normal} functor if it
is has the form 
\[
N = \oplus_n (\otimes_{m_1}(-),\ldots,\otimes_{m_n}(-)).
\]

\begin{lemma}\label{lem:complete1}Every functor $G$ is naturally isomorphic to a normal
functor $N_G$.

\begin{proof}[Sketch proof]
Use induction on the structure of $G$;  the required isomorphism is
constructed from the distributivity $ A\otimes (B\oplus C) \iso
(A\otimes B)\oplus (A\otimes C)$.
\end{proof}

\begin{proof}[Long proof]
We construct $N_G$ and a natural isomorphism $d_G$ simultaneously, by
recursion on the structure of $G$.  There are two principal cases.

If $G = G_1(-)\oplus G_2(-)$ then, by induction,  we have natural
isomorphisms $d_1 : G_1 \Rightarrow N_{G_1}$ and $d_2 : G_2
\Rightarrow N_{G_2}$. Then $N_{G_1}\oplus N_{G_2}$ is a normal
functor, and $d_1\oplus d_2$ is the required natural isomorphism.

If $G = G_1(-)\otimes G_2(-)$ then we have natural isomorphisms 
$d_1 : G_1 \Rightarrow N_{G_1}$ and $d_2 : G_2 \Rightarrow N_{G_2}$.
Since $N_{G_1}$ is normal it has the form $\bigoplus_i A_i$,
where each $A_i$ is multi-ary tensor product; 
similarly $N_{G_2} = \bigoplus_j B_j$.  Hence we have a natural
isomorphism 
\[
        d = \langle \langle \pi_i \otimes 1 \rangle_i \otimes \pi_j
\rangle_j : N_{G_1}\otimes N_{G_2} \Rightarrow \bigoplus_{ij}
A_i\otimes B_j
\]
to a normal functor, which we take to be $N_G$.  The composition
$d\circ(d_1\otimes d_2)$ is the required map $G \Rightarrow N_G$.
\end{proof}
\end{lemma}

\begin{lemma}\label{lem:complete2}
Let $N_F,N_G$ be normal functors.  For each arrow $f:N_F\overline{A}
\to N_G\overline{B}$ there exist 
maps $g,h,f_1,\ldots,f_n$ such that 
\begin{diagram}
N_F\overline{A} & \rTo^f & N_G\overline{B} \\
\dTo^g && \uTo_h\\
\bigoplus_i A_i & \rTo_{\qquad\bigoplus_i f_i\qquad} & \bigoplus_{i}  B_j
\end{diagram}
commutes, where the $A_i,B_j$ are multi-ary tensors of literals.
\begin{proof}
We will construct the required maps by recursion on the structure of
$f$.  There are three cases.

\textbf{Case 1.}
Suppose $A = A_1\oplus A_2$, and $B = B_1\oplus B_2$.  Then $f$ has a
matrix representation $\left(\begin{array}{cc}f_1&f_2\\f_3&f_4\end{array}\right)$.
We reconstruct $f$ as
\begin{diagram}
A_1\oplus A_2 & \rTo^f & B_1 \oplus B_2 \\
\dTo^{\Delta_{A_1}\oplus\Delta_{A_2}} && \uTo_{\nabla_{B_1\oplus B_2}}\\
A_1\oplus A_1 \oplus A_2\oplus A_2 & 
\rTo_{\qquad f_1\oplus f_2 \oplus f_3\oplus f_4 \qquad} & 
B_1\oplus B_2 \oplus B_1\oplus B_2
\end{diagram}
and recurse on each  $f_i$.

\textbf{Case 2.}
Suppose that $A = A_1\oplus A_2$, but $B$ is not a biproduct of two
other objects.  Since the biproduct structure is freely constructed it
is guaranteed that $f = [f_1,f_2]: A_1\oplus A_2 \to B$.  This is
reconstructed as
\begin{diagram}
A_1\oplus A_2 & \rTo^{f_1\oplus f_2} & B\oplus B & \rTo^{\nabla_B} & B.
\end{diagram}

\textbf{Case 3.} If $B = B_1\oplus B_2$ but $A$ is not a
biproduct the treatment is dual to that of case 2.
\end{proof}
\end{lemma}

Now consider the $f_i$ constructed above.  Each one has the form 
\[
f_i : \otimes_n\overline{A}\to \otimes_m\overline{B}.
\]
Suppose that \textbf{0} is a component of $\overline{A}$.  In that
case $\otimes_n\overline{A} =  \textbf{0}$ and hence $f_i$ is completely
determined; a similar situation applies to $\overline{B}$.  Let us
suppose then, that \textbf{0} does not occur in either $\overline{A}$
or $\overline{B}$.  

Since the hom sets of $F\catA$ form a (freely generated) commutative
monoid, $f$ is a finite sum of non-zero arrows, $f_i = \sum_j
f_{ij}$, or else is zero itself.  Furthermore, every $f_{ij}$ must be
an arrow of the subcategory $F_{\mathrm{SCC}}\catA$, that is the free strongly
compact closed category upon $\catA$.  (Here we are relying on Proposition~\ref{FCCfaithembedprop}).
At this point we appeal to 
a theorem of \cite{KelLap:comcl:1980, Calco}:

\begin{theorem*}
Each arrow $f:A\to B$ of the free (strongly) compact closed category on a category \catA is completely
described by the following data:
\begin{enumerate}
\item An involution $\theta$ on the atoms of $A^*\otimes B$;
\item A functor $p: \theta \to \catA$ agreeing with $\theta$ on
objects (i.e. a labelling of $\theta$ with arrows of \catA.);
\item A multiset $L$ of loops from \catA.
\end{enumerate}
\end{theorem*}

\begin{lemma}\label{scclemma}
Suppose that $\catA$ has an identity on objects, contravariant, involutive functor $()^{\dagger}$. Then $F_{\mathrm{SCC}}\catA$ is strongly compact closed.
\begin{proof}
It suffices to show how to extend $()^{\dagger}$ to $F_{\mathrm{SCC}}\catA$. Using the description of morphisms in $F_{\mathrm{SCC}}\catA$ given in the previous theorem, we define
\[ (\theta , p, L)^{\dagger} = (\theta^{-1}, ()^{\dagger} \circ p \circ ()^{-1}, L^{\dagger}) . \]
\end{proof}
\end{lemma}

\begin{lemma}
For each arrow $f$ in $F_{\mathrm{SCC}} \catA$ there is a proof-net $\nu$ such
that $\denote{\nu} = \name{f}$.
\begin{proof}
By Kelly-Laplaza  $f \approx (\theta, p, L)$. The involution $\theta$
specifies the axiom links, labelled as per the functor $p$.  We add
tensor links to join up all the conclusions which are subformulae of
$A$, and likewise $B$.  For each
loop $h:A\to A$ in $L$,  an $h$-axiom link is added; the loop is closed up
with an identity cut.  Since $\name{s\bullet f} = s\bullet\name{f}$
this suffices.
\end{proof}
\end{lemma}

\begin{figure}[tbp]
\topfigrule
\figureline
\begin{center}
  \includegraphics{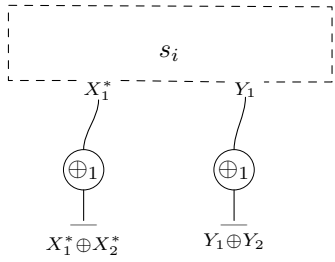}
\end{center}  
\caption{Proof-net for lemma~\ref{lem:plusandnames}} \label{fig:pnforlemma}
\figureend
\end{figure}

\begin{lemma}\label{lem:plusandnames}
  Given $f_1:X_1\to Y_1$ and $f_2:X_2\to Y_2$, if there exist
  proof-nets $\pi_1,\pi_2$ such that   $\denote{\pi} = \name{f}$ and
  $\denote{\pi_2}$ then there exists $\pi$ such that $\denote{\pi} =
  \name{f_1\oplus f_2}$.
\end{lemma}
\begin{proof}
  First note that 
  \[
  f_1 \oplus f_2 = (q_1 \circ f_1 \circ p_1) + (q_2  \circ f_2 \circ p_2),
  \]
  where $p_i,q_j$ are the biproduct projections and injections.  Hence
  \begin{eqnarray*}
    \name{f_1\oplus f_2} & = & 
    (     
      \id{X_1^*\oplus X_2^*}
      \otimes
     ((q_1 \circ f_1 \circ p_1) + (q_2 \circ f_2 \circ p_2)) 
    ) 
    \circ \eta_{X_1\oplus X_2}
\\&=&
    (( \id{X_1^*\oplus X_2^*} \otimes (q_1 \circ f_1 \circ p_1))
    \circ \eta_{X_1\oplus X_2}) 
    + 
    ((\id{X_1^*\oplus X_2^*} \otimes (q_1 \circ f_1 \circ p_2))
    \circ \eta_{X_1\oplus X_2}) 
\\&=&    
    (( p_1^* \otimes (q_1 \circ f_1))
    \circ \eta_{X_1} ) 
    +
    (( p_2^* \otimes (q_2 \circ f_2))
    \circ \eta_{X_2}) 
\\&=&
    (( p_1^* \otimes (q_1 \circ f_1))
    \circ \eta_{X_1} ) 
    +
    (( p_2^* \otimes (q_2 \circ f_2))
    \circ \eta_{X_2}) 
\\&=&
    ((q_1\otimes q_1)\circ (\id{X^*_1}\otimes f_1) \circ \eta_{X_1}) 
    + ((q_2\otimes q_2)\circ (\id{X^*_2}\otimes f_2) \circ \eta_{X_2})
\\&=&
    ((q_1\otimes q_1)\circ \name{f_1}) 
    + ((q_2\otimes q_2)\circ \name{f_2}).
\\&=&
    ((q_1\otimes q_1)\circ \denote{\pi_1}) 
    + ((q_2\otimes q_2)\circ \denote{\pi_2}).
  \end{eqnarray*}
If $\pi_1$ has slices $s_i$ then
\[
(q_1\otimes q_1)\circ \denote{\pi_1} = 
(q_1\otimes q_1)\circ \sum_i\denote{s_i} = 
\sum_i(q_1\otimes q_1)\circ\denote{s_i}.
\]
hence we require a slice $s'_i$ such that $\denote{s'_i} = (q_1\otimes
q_1)\circ\denote{s_i}$.  We assume that $s_i$ has conclusions $X_1^*$ and
$Y_1$;  if it does not, then necessarily its conclusions differ from this
only by the arrangement of the tensor links, and since these have no
impact on the denotation we can rearrange them as needed.  Then the
required slice is shown in Fig.~\ref{fig:pnforlemma}.  The required
proof-net $\pi$ is formed by combining all the $s'_i$ into a single
proof-net, along with similarly constructed slices for $f_2$.  By its
construction we have $\denote{\pi} = \name{f_1\oplus f_2}$.
\end{proof}
\end{document}